\documentclass[12pt]{article}
\usepackage[colorlinks]{hyperref}
\newcommand{\ba}{\begin{array}}
\newcommand{\ea}{\end{array}}
\newcommand{\be}{\begin{equation}}
\newcommand{\ee}{\end{equation}}
\newcommand{\nn}{\nonumber}
\newcommand{\bea}{\begin{eqnarray}}
\newcommand{\ena}{\end{eqnarray}}
\newcommand{\beas}{\begin{eqnarray*}}
\newcommand{\enas}{\end{eqnarray*}}
\newcommand{\mb}{\mbox}

\addtocounter{section}{0}
\begin{document}


\begin{center}
{\LARGE On the solutions of the Yang-Baxter equations with general
inhomogeneous eight-vertex $R$-matrix:}
 {\large Relations with
 Zamolodchikov's tetrahedral algebra}
\end{center}

\begin{center}
{\bf  Sh. Khachatryan\footnote{e-mail:{\sl shah@mail.yerphi.am}},
A. Sedrakyan\footnote{e-mail:{\sl sedrak@nbi.dk}} }\\ Yerevan
Physics Institute, Alikhanian Br. str. 2, Yerevan 36, Armenia
\end{center}

\vspace{1.5 cm}
\begin{center}
{\LARGE
 Abstract}
\end{center}

We present  most general one-parametric solutions of the
Yang-Baxter equations (YBE) for one spectral parameter dependent
$R_{ij}(u)$-matrices of the six- and eight-vertex models, where
the only constraint is the particle number conservation by mod(2).
A complete classification of the solutions is performed. We have
obtained also  two spectral parameter dependent particular
solutions $R_{ij}(u,v)$ of YBE.  The
 application of the non-homogeneous solutions to construction
 of Zamolodchikov's tetrahedral algebra is discussed.



\section{Introduction}

Yang-Baxter equations (YBE) play an important role in theory of
two-dimensional integrable models
\cite{Y,Baxt78,FT,Baxt,FTS,Z,Z1}. They ensure the presence of
 infinite amount of conservation laws
making model integrable. 
 Integrable models arise in many areas of physics (statistical
mechanics, high energy physics, condensed matter physics, string
theory, atomic and molecular physics), as well as the mathematical
methods and tools developed in the theory of integrable systems
have broad applications in different branches of  modern physics
(e.g. see the works \cite{FT,Baxt,FTS,Ons,QISM,conform,FR} and
references therein). A major element in the theory of
two-dimensional integrable
 models is  $R$-matrix, which should satisfy
YBE. $R$-matrices are classified by degrees of freedom of the
chain sites, where they are acting, and by the symmetries of the
model. Usually minimal, simplest solutions of YBE are enough to
characterize the model. The $R$-matrices of the anisotropic
Heisenberg, Hubbard, Uimin-Lai-Sutherland models are just simplest
solutions of corresponding YBE and they carry most important
properties of the corresponding classes. However, one can expect,
that there can be a generalization of the model by keeping
integrability, which, nevertheless, will correspond to another
physical situation.
In this context a substantial question is rising about the
properties of the statistical (Boltzmann) weights in two
dimensional statistical physics \cite{Baxt,FTS,Z1}, or the
scattering $R$-matrices in the $1+1$ quantum field theories
\cite{Y,Z}, which maintain the integrability.
 The general form of the spectral
parameter dependent YBE
is the following
\be
R_{12}(u,v)R_{13}(u,w)R_{23}(v,w)=R_{23}(v,w)R_{13}(u,w)R_{12}(u,v).\label{ybe}\ee
 In
this paper we consider the one parametric class of $R$-matrices,
$R(u,w)=R(u-w)$. $R_{12}$ matrix acts on the direct product of two
states, $|i_1\rangle|i_2\rangle$. The variables  $i_k$ describe
the degrees of  states at the lattice sites (in context of
$1$-dimensional quantum chain models or $2$-dimensional models in
classical statistical mechanics) or the states of the scattering
particles (in context of $(1+1)$-dimensional quantum scattering
theory). Let us constrain  ourselves to the case of two
dimensional quantum spaces at the sites, namely when  $i_k$ can
have only two values (spin-$\frac{1}{2}$ system).

Usually some characteristics of the $R$-matrix of the model are
fixed on the basis of symmetry properties of the underlying
physical problem. Here we fix only the structure of
 $R$-matrices, i.e the non-vanishing elements positions. The only symmetries, which is taken
 into account, is the "particle number"
conservation by mod(2) ($\mathcal{Z}_2$ grading symmetry). In matrix
representation it means
\bea R_{ij}^{kr}\neq 0 \qquad \mbox{if}\qquad i+j+k+r=0 (mod\;
2).\label{rij}\ena
If $i,j...$ take only two values $0,\; 1$, then $R$ has a following $4\times
4$ matrix form
 \bea \label{rg}R(u-v)=\left(\ba{cccc}
R_{00}^{00}&0&0&R_{00}^{11}\\
0&R_{01}^{01}&R_{01}^{10}&0\\
0&R_{10}^{01}&R_{10}^{10}&0\\
R_{11}^{00}&0&0&R_{11}^{11}\\
\ea\right)
,\ena
corresponding to the $R$-matrix of the eight-vertex model.

 There are well known 
symmetric solutions of the YBE
 (\ref{ybe}) with the form of (\ref{rg}), namely the
elliptic $R$-matrix of the $XYZ$ model \cite{Baxt}, the solutions
with free fermionic property \cite{WuChan} (see also $R$-matrix
for $2d$ Ising Model (IM) in \cite{ShAS} and the three-parametric
solutions to the YBE in \cite{BS}). But, although the YBE with
 the  eight-vertex model's type $4\times 4$ $R$-matrices are
investigated and  crucial solutions are obtained, however there is
an open question remained here about the completeness of the
solutions, as the authors usually take physically motivated
symmetric matrices and the existed classification  of the
solutions reflects this fact. Here we present a full
classification and the complete list of the solutions with the
given form and also we show, that the YBE themselves are putting
restrictions on the elements of $R(u)$, dictating some symmetry
relations on them (namely, relations among the elements
$R^{ij}_{kr}$ and $R^{1-i\; 1-j}_{1-k\; 1-r}$). Although we
consider only the $4\times 4$-matrices, but obviously such kind of
behavior is valid also for the high dimensional matrices.

 In the  Section 2 we present consistency conditions
 for six- and eight-vertex type one-parametric $R$-matrices to be  solutions to YBE,
and  a complete list of the nontrivial solutions.
 For both cases  the compatibility
conditions, expressed by the formulas (\ref{sc}, \ref{sb},
 \ref{summarize1}) and
(\ref{ast12}, \ref{summarizeg}), 
 can be proclaimed 
 by means of two statements: {\it there are  symmetric relations
 followed from YBE
between the matrix elements and also the matrix elements 
obey expected homogeneous equations of the
second order}. 
 In the Section 3 the general
parametrization for the general eight-vertex type solutions is
given by means of trigonometric and elliptic functions. The
solutions of the homogeneous eight-vertex model obtained in
\cite{Baxt} correspond to the case presented in the subsection
3.1. In the subsections 3.2 and 3.3 we present the general
nonhomogeneous solutions. Next section is devoted to the
description of the Hamiltonian operators of the corresponding 1d
quantum spin-chain models.

In the Section 5, for completeness, we give the investigation of
$R$-matrices,
which we have excluded in the first two sections, since they
contain more vanishing matrix elements than it is presented in
(\ref{rg}), or have coinciding matrix elements, which in many
cases leads to constant solutions.  Investigations of the constant
$R$-matrices are performed in the papers \cite{Hieta,Hlavati}. We
 consider the spectral parameter dependent solutions, and it
turns out, that along with the solutions, which can be obtained
from the matrices considered in the Sections 2,3, after taking the
appropriate limits, there are also independent solutions.

The  Section 6 is devoted to the analysis of the tetrahedral
Zamolodchikov's algebra \cite{Korep}
 with the inhomogeneous
$R$-matrices.

\section{Investigation of the solutions to the YBE}
\addtocounter{section}{0}\setcounter{equation}{0}

 In the consideration below we analyze Yang-Baxter equations for spin-$\frac{1}{2}$ systems in its most general form.
 These equations are well known, however we consider it is worthy to give
 one more time all the relations obviously for performing
 detailed analysis and for giving an exhaustive answer to the question
 what are the all  one-spectral parameter dependent solutions
 in the form of  (\ref{rg})
 to the YBE (\ref{ybe}).

We use the following notations  for the matrix elements
%
 \bea \nn
&R^{00}_{00}=a_1(u),\; R^{11}_{11}=a_2(u),\;
R^{01}_{01}=b_1(u),\;R^{10}_{10}=b_2(u),\;&\\
 &R^{10}_{01}=c_1(u),\; R^{01}_{10}=c_2(u),\;
 R^{11}_{00}=d_1(u),\; R^{00}_{11}=d_2(u).&
\label{rx}\ena
\subsection{XXZ type $R$-matrices}


 \setcounter{equation}{0}

In the beginning let us  briefly start with the case, when no
creation or annihilation of the pairs may occur, if the $R$-matrix
is assumed as particle's scattering matrix. It means that all the
non-vanishing elements have the
 property $i+j=k+r$, together with (\ref{rij}).
When the dimension of the states, where the $R$-matrix is acting,
is two, it has  form of the $XXZ$ model's
 $R$-matrix, i.e. $d_1=d_2=0$ in (\ref{rx}). As we shall see later,
 the consideration of the general case ($d_i\neq 0$) does not reflect all the peculiarities
 of this particular case (taken in the appropriate limit $d_i\to
 0$).
%

The YB equations in the matrix elements notations are given as
follows
\be \sum_{j_1,j_2,j_3} R_{i_1 i_2}^{j_1 j_2}(u)R_{j_1 i_3}^{k_1
j_3}(u+w)R_{j_2 j_3}^{k_2 k_3}(w)=\sum_{j_1,j_2,j_3} R_{i_2
i_3}^{j_2 j_3}(w)R_{i_1 j_3}^{j_1 k_3}(u+w)R_{j_1 j_2}^{k_1
k_2}(u). \label{ybee}\ee
%
From the simple equations existing in (\ref{ybee}) of this kind
\be c_1(u)c_1(w) c_2(u+w) = c_1(u+w) c_1(u) c_1(w), \label{c}\ee
it follows that $c_1(u)/c_2(u)$ is an exponential function $e^{u
\alpha}$, where $\alpha$ is an arbitrary number. In the remaining
$12$ equations one can distinguishes $6$ pairs of the equations,
such that in the each pair one of the equations will coincide with
another, after using the relation (\ref{c}). Taking into account
this fact we choose the following six independent equations.
{\footnotesize\bea \nn a_1(u+w) b_1(u) c_1(w) - a_1(u) b_1(u+w)
c_1(w) + b_1(w) c_1(u+w) c_2(u) = 0,\\\nn a_1(u) a_1(w) c_1(u+w) -
b_1(w) b_2(u) c_1(u+w) -a_1(u+w) c_1(u) c_1(w) = 0,\\\label{ybeh}
 a_1(w) b_2(u+w) c_1(u) - a(u+w) b_2(w) c_1(u) -
b_2(u) c_1(u+w) c_2(w) = 0,\\\nn
 a_2(w) b_1(u+w) c_1(u) - a_2(u+w) b_1(w) c_1(u) - b_1(u) c_1(u+w) c_2(w)
= 0,\\\nn
 a_2(u) a_2(w) c_1(u+w) - b_1(u) b_2(w) c_1(u+w) - a_2(u+w)
c_1(u) c_1(w) = 0,\\\nn
 a_2(u+w) b_2(u) c_1(w) - a_2(u) b_2(u+w)
c_1(w) + b_2(w) c_1(u+w) c_2(u) = 0. \ena}
The equations above are linear and homogeneous with respect of the
functions $a_1(w)$, $b_1(w)$, $c_1(w)$, $a_2(w)$, $b_2(w)$,
$c_2(w)$ and so have  non-zero solutions, if the determinant of
the matrix of the corresponding coefficients vanishes. The
determinant reads as
\bea \label{re} & \Big(a_1(u) a_2(u+w) b_1(u+w) b_2(u) - a_1(u+w)
a_2(u) b_1(u) b_2(u+w)\Big)&\\\nonumber &\times\Big(a_1(u) a_2(u)
+ b_1(u) b_2(u) - c_1(u) c_2(u)\Big)c_1(u)c_2(u) c_1(u+w)^4.&\ena
We are excluding the cases $c_{1,2}(u)=0$ by  now 
 and will observe that situation in the Section 5, which
includes the exceptional cases.

\paragraph{1. $^\ast$}
The vanishing of the first bracket in the determinant's expression
means
\bea \frac{a_1(u) b_2(u)}{a_2(u) b_1(u)}={\rm{constant}}\equiv
b_0.\label{r1}\ena
Using this relation, from the first five equations presented above
(\ref{ybeh}), we derive two possible relations. One of them is
$a_1(u)=b_1(u)\sqrt{\frac{-1}{b_0}}$, which leads to 
the following subsequent equations:  $a_1(u)=a_2(u)\quad  \&\quad
b_1(u)c_1(u+w)c_2(w)=0$ (for such solutions see Section 5). The
other relation implies
\bea 
%
\frac{a_1(u)a_2(u)+b_1(u)b_2(u)-c_1(u)c_2(u)}{a_1(u)b_2(u)}=\rm{constant}\equiv
\Delta.\label{cx}\ena
In this case (\ref{cx})
the remaining equations give the following additional
requirements:
\bea
 b_1(u)=b_0 \;b_2(u)\quad \& \quad a_1(u)=a_2(u).
\ena
%
 %

\paragraph{2.$^{\ast\ast}$} The vanishing of the second
bracket in the expression (\ref{re}) means that the following
relation holds
\bea a_1(u) a_2(u) + b_1(u) b_2(u) - c_1(u)
c_2(u)=0.\label{r2}\ena
If $a_1(u)=a_2(u)$, then from the equations (\ref{ybeh}) we  find
$b_1(u)= b_2(u) \;\textsf{constant}$, and this case becomes
equivalent to the first discussed case (i.e. \ref{r1}, \ref{cx})
with $\Delta=0$.

 Let us observe now the situation, when $a_1(u)\neq
a_2(u)$. From the first, third and forth equations in
(\ref{ybeh}), taking into account that the change
$u\leftrightarrow v$ in the equations gives permissible and
equivalent equations, we find relations
\bea \frac{a_2(u)-a_1(u)}{b_1(u)}= \bar{\;\Delta}\quad {\rm{and}}
\quad \frac{b_1(u)}{b_2(u)}=b_0,\label{aa}\ena
where $\bar{\Delta}$ and $b_0$ are constants. By means of the
above relations (\ref{aa}) the constraint (\ref{r2}) takes the
following form
\bea a_1(u)^2 + b_1(u)^2 b_0 - c_1(u) c_2(u)=\bar{\Delta} a_1(u)
b_1(u),\label{cx1}\ena
reminding the relation (\ref{cx}).

 The possibilities $\mathbf{1}$ and $\mathbf{2}$
  listed above are
 overlapping
 when $\Delta=0$ and $\bar{\Delta}=0$.

\paragraph{Let us summarize.} The general solutions of the YBE with
$R$-matrix (\ref{rx}) have the following properties (we are
omitting the arguments $(u)$ of the functions $f_i(u)$, $f=a,b,c$
for simplicity)
\bea
&c_1/c_2=e^{\alpha u},&\label{sc}\\
&b_1/b_2=\textit{b}_0&\label{sb}\\
&\{a_1=a_2 \; \& \; a_2a_1-c_1c_2+b_1b_2=a_1b_1\Delta\}^* \;
\mb{or}\; \{a_2-a_1=\bar{\Delta} b_1 \;\&\;
a_2a_1=c_1c_2-b_1b_2\}^{**}.&\nonumber \\\label{summarize1}\ena
Here $\alpha,\Delta,\bar{\Delta},\textit{b}_0$ are arbitrary
constants (they don't depend from the spectral parameter $u$).

The most general solutions for two cases can be presented by means
of  the trigonometric 
parametrization 
(it is enough to choose $a_1(u)=\sin{[u+u_0]}$ and
$c_1(u)=e^{\frac{\alpha u}{2}}\sin{[u_0]}$, as for the XXZ
$R$-matrix, then the expressions for the functions $b_i(u)$,
$c_2(u)$ and $a_2(u)$ will follow from (\ref{sc}, \ref{sb},
\ref{summarize1}))
 \be R^{\ast/\ast\ast}(u)=\left( \ba{cccc} \sin{[u+u_0]}&0&0&0\\
0&\sin{[u]}\sqrt{b_0} &e^{\frac{\alpha u}{2}}\sin{[u_0]}&0\\
0&e^{-\frac{\alpha u}{2}} \sin{[u_0]}&\sin{[u]}\frac{1}{\sqrt{b_0}}&0\\
0&0&0&\sin{[u_0]}\pm u)
 \ea \right).\label{rxxzi}\ee

 Any other parametrization can be obtained from this one either by the
 redefinitions of the arguments or by basis change.  The rational solutions
 correspond to the limit $\lim_{h\to 0}{R(h u)/\sin{[hu]}}$. 

The first case ${ }^*$ in (\ref{summarize1}) corresponds to the
ordinary XXZ model (intertwiner matrix of the $sl_q(2)$ algebra at
general $q$). The second logarithmic derivative of the transfer
matrix constructed by this matrix gives
 the Hamiltonian $H$ of the anisotropic Heisenberg magnetic with
 anisotropy parameter $\Delta/2=\cos{[u_0]}$.

If we represent $R$ by means of the tensor products of the Pauli
$\sigma$ matrices, then $H$ becomes (we take $b_0=1, \alpha=0$)
\bea & H^{*}=\sum_k
\Big(\sigma_1(k)\sigma_1(k+1)+\sigma_2(k)\sigma_2(k+1)+
\frac{\Delta}{2}\sigma_3(k)\sigma_3(k+1)\Big).& \ena

The second case ($^{\star\star}$)
 corresponds to the $XX$-model in the transverse magnetic field
 $h=\bar{\Delta}/2=\cos{[u_0]}$ (intertwiner matrix of the $sl_q(2)$ matrix at $q=i$, nilpotent \cite{GRS} and cyclic
 irreps \cite{shkh}).
\bea & H^{**}=\sum_k
\Big(\sigma_1(k)\sigma_1(k+1)+\sigma_2(k)\sigma_2(k+1)+
\bar{\Delta}\sigma_3(k)\Big).& \ena

We see, that the only  significantly different solution from
 the known $XXZ$-model's $R$-matrix, suggests
 $R_{11}^{11}\neq R_{00}^{00}$. The appearance
of the function $e^{\alpha u}$ can be addressed to the change of
the basis vectors of the definition spaces. 

\subsection{YBE with extended XYZ type $R$-matrices} 

If one allow the creation or annihilation of the pairs in the
scattering matrix (eight-vertex model), we must deal with the more
general form of the $R$-matrix (\ref{rx}).
The simplest equations among the YBE (\ref{ybee}) are
\bea c_1(u)c_1(w) c_2(u+w) - c_1(u+w) c_2(u) c_2(w)=0,\label{eq1}\\
-c_1(u+w) d_1(w) d_2(u) + c_2(u+w) d_1(u) d_2(w)=0,\label{eq2}\\
 c_1(u) d_1(w) d_2(u+w) -
c_2(u) d_1(u+w) d_2(w)=0.\label{eq3}\ena
Let $c_i\neq 0$. The equation (\ref{eq1}) leads to
$c_1(u)=c_2(u)e^{k u}$, with arbitrary constant $k$. Placing this
relation into the equations (\ref{eq2}, \ref{eq3}), we come to
$k=0$ ({\it note}, that  when $d_i=0$, $k$ is arbitrary, see
subsection 2.1). Then $d_2(u)=d_0 d_1(u)$, $d_0$ is an arbitrary
constant. So we have
\bea c_1(u)=c_2(u),\qquad d_2(u)=d_0 d_1(u). \label{rcd}\ena

From the analysis of the previous case with matrix (\ref{rx}) we
learn, that there are two different non trivial restrictions on
the solutions to YBE. One corresponds to the
case $a_1(u)=a_2(u)$, 
 the second one to $a_1(u)\neq a_2(u)$
.

\paragraph{1. $^\ast$} Let us at first consider the case
$a_1(u)=a_2(u)$.

Comparing the following equations from the set of the YBE
(\ref{ybee})
\bea a_1(w) (c_1(u\!+\!w) d_1(u)\! -\!c_1(u) d_1(u\!+\!w))\! + \!
(a_1(u) a_1(u\!+\!w)
 \!-\! b_1(u) b_1(u\!+\!w)) d_1(w) = 0,\\
 a_1(w) (c_1(u\!+\!w) d_1(u)\! -\! c_1(u) d_1(u\!+\!w))\! +\! (a_1(u) a_1(u\!+\!w)
\!-\! b_2(u) b_2(u\!+\!w)) d_1(w) = 0,\ena
we immediately find
\be b_2(u)=b_0\; b_1(u), \qquad b_0^2=1. \label{rb}\ee
{\it Note}, that  when $d_i=0$, $b_0$ is an arbitrary constant.

Taking into account (\ref{rcd}) and (\ref{rb}),  there are only
six independent equations in the YBE (quite similarly to the XYZ
case, discussed in the work \cite{Baxt}).
\bea \nn &a(w) c(u+w) d(u) - a(w) c(u) d(u+w) + a(u) a(u+w) d(w) - b(u) b(u+w) d(w) = 0;&\\
\nn & -b(w) c(u) c(u+w) - a(u+w) b(u) c(w) + a(u) b(u+w) c(w) + b_0 d_0b(w)d(u) d(u+w) = 0;&\\
\nn
&a(u) a(w) c(u+w) - b_0 b(u) b(w) c(u+w) - a(u+w) c(u) c(w) + d_0 a(u+w)  d(u) d(w) = 0;&\\
 \nn &a(w) b_0
b(u+w) c(u) - b_0 (a(u+w) b(w) c(u) + b(u) c(u+w) c(w)) + d_0 b(u)  d(u+w) d(w) = 0;&\\
\nn &b_0 b(u+w) c(w) d(u)
-b_0 a(w) b(u) d(u+w) + a(u) b(w) d(u+w) - b(u+w) c(u) d(w) = 0; &\\
\nn &-a(u+w) a(w) d(u) + b(u+w) b(w) d(u) + a(u) c(w) d(u+w) -
a(u) c(u+w) d(w) = 0. &\\\label{ybez}\ena
In the given equations we are omitting  the index $1$, taking
$f_1(u)\equiv f(u),\; f=a,b,c,d$. The consistency condition of the
last four equations is
\bea \label{cz}&\Big(a(u\!+\!w) b(u\!+\!w) c(u) d(u)\! -\! b_0
a(u) b(u) c(u\!+\!w) d(u\!+\!w)\Big)\times &\\\nn &\Big((a^2(u)
\!-\! b^2(u))(c^2(u\!+\!w) \!-\! b_0 d_0 d^2(u\!+\!w))\! +\!
(a^2(u\!+\!w)\! -\! b^2(u\!+\!w))(b_0 c^2(u) \!-\! d_0
d^2(u))\Big)=0.&\ena
As we see there are two cases that one must observe.
\paragraph{1.1} If the first bracket in (\ref{cz}) vanishes,
\bea a(u\!+\!w) b(u\!+\!w) c(u) d(u)\! -\! b_0 a(u) b(u)
c(u\!+\!w) d(u\!+\!w)=0,\label{cz1} \ena
the following relations are true
\bea  \frac{a(u) b(u)}{c(u) d(u)}={\rm{constant}} ,\qquad
b_0=1.\label{cz10} \ena
This is identical to the case, considered in \cite{Baxt} with
homogeneous and symmetric  $R$-matrix. As it is known, the
remaining equations give the constraint
\bea
\frac{a^2(u)+b^2(u)-c^2(u)-d^2(u)}{a(u)b(u)}={\rm{constant}}.\label{czg}\ena

This is the well observed case (see \cite{Baxt}) of the XYZ
model's $R$-matrix
 (which will be given in a precise form in the next section) satisfying this
constraint.

\paragraph{1.2} The vanishing of the second bracket in (\ref{cz})
\bea \left(a^2(u) \!-\! b^2(u)\right)\left(c^2(u\!+\!w) \!-\! b_0
d_0 d^2(u\!+\!w)\right)\! +\! \left(a^2(u\!+\!w)\! -\!
b^2(u\!+\!w)\right)\left(b_0 c^2(u) \!-\! d_0
d^2(u)\right)=0,\nn
\ena
gives the relations
\bea  \frac{a^2(u) \!-\! b^2(u)}{b_0 c^2(u) \!-\! d_0
d^2(u)}={\rm{constant}}, \qquad b_0=-1.\label{cz20} \ena
%

The constant in (\ref{cz20}) can be fixed from the analysis of the
remaining equations, or by a rather simple way. Let us fix the
variable $w$ to be $0$, then from the set of the YBE we can see,
that non-trivial ($a(u)\neq \pm b(u),\;\;\; d(u)\neq \pm c(u)$)
solutions demand $a(0)=c(0)$ and $b(0)=d(0)=0$. This fixes the
constant to be $-1$. So
\bea  a^2(u) - b^2(u)- c^2(u) - d_0 d^2(u)=0.\label{cz20x} \ena
%

 A particular case of (\ref{cz20}) is (below $x_1,\; x_2$ are constant numbers)
\bea  a^2(u) \!-\! b^2(u)=x_1, \qquad c^2(u) \!+\! d_0 d^2(u)=x_2,
\qquad b_0=-1.\label{cz2012} \ena
Analyzing the equations (\ref{ybez}) we find that $x_1=0,\;x_2=0$.
Such exceptional cases will be discussed ($f_i(u)=0$ or $f(u)=\pm
g(u)$, $f,\;g=a,b,c,d$) in the Section 5. %
%
%

\paragraph{2.$^{\ast \ast}$} Now let us consider the case $a_1(u)\neq a_2(u)$.

The YBE now contain twelve independent equations.
\bea\nn a_1(u) b_1(u\!+\!w) c_1(w) \!-\!b_1(w) c_1(u) c_1(u\!+\!w)
\!-\! a_1(u\!+\!w) b_1(u) c_1(w) \!+\! d_0 b_2(w) d_1(u)
d_1(u\!+\!w) = 0,
\\\nn a_1(u) a_1(w) c_1(u\!+\!w) \!-\! b_1(w) b_2(u) c_1(u\!+\!w) \!-\! a_1(u\!+\!w) c_1(u) c_1(w) \!+\! a_2(u\!+\!w) d_0 d_1(u) d_1(w) = 0,\\\nn a_1(w)
b_2(u\!+\!w) c_1(u) \!-\! a_1(u\!+\!w) b_2(w) c_1(u) \!-\! b_2(u) c_1(u\!+\!w) c_1(w) \!+\! d_0 b_1(u) d_1(u\!+\!w) d_1(w) = 0,\\
\nn
a_2(u\!+\!w) b_1(w) c_1(u) \!+\! b_1(u) c_1(u\!+\!w) c_1(w) \!-\! d_0 b_2(u) d_1(u\!+\!w) d_1(w)\!-\!a_2(w) b_1(u\!+\!w) c_1(u) = 0,\\
\nn b_1(u) b_2(w) c_1(u\!+\!w) \!+\! a_2(u\!+\!w) c_1(u)
c_1(w)\!-\!a_2(u) a_2(w) c_1(u\!+\!w) \!-\!d_0  a_1(u\!+\!w)
d_1(u) d_1(w) = 0,\\
\nn a_2(u) b_2(u\!+\!w) c_1(w) \!+\! d_0 b_1(w)  d_1(u) d_1(u\!+\!w) \!-\!b_2(w) c_1(u) c_1(u\!+\!w) \!-\! a_2(u\!+\!w) b_2(u) c_1(w) = 0, \\
\nn a_2(w) c_1(u\!+\!w) d_1(u) \!-\! a_1(w) c_1(u) d_1(u\!+\!w) \!+\! a_1(u) a_1(u\!+\!w) d_1(w) \!-\! b_1(u) b_1(u\!+\!w) d_1(w) = 0,\\
\nn b_2(u\!+\!w) c_1(w) d_1(u) \!+\! a_1(u) b_1(w) d_1(u\!+\!w)
\!-\! a_1(w) b_2(u) d_1(u\!+\!w) \!-\! b_1(u\!+\!w) c_1(u) d_1(w)
= 0,\\ \nn   b_2(u\!+\!w) b_2(w) d_1(u) \!+\! a_1(u) c_1(w) d_1(u\!+\!w) \!-\! a_2(u) c_1(u\!+\!w) d_1(w) \!-\!a_1(u\!+\!w) a_1(w) d_1(u)= 0,\\
\nn a_2(u\!+\!w) a_2(w) d_1(u) \!-\! b_1(u\!+\!w) b_1(w) d_1(u) \!-\! a_2(u) c_1(w) d_1(u\!+\!w) \!+\! a_1(u) c_1(u\!+\!w) d_1(w) = 0,\\
\nn a_2(w) b_1(u) d_1(u\!+\!w) \!-\! a_2(u) b_2(w) d_1(u\!+\!w) \!+\! b_2(u\!+\!w) c_1(u) d_1(w)\!-\!b_1(u\!+\!w) c_1(w) d_1(u)  = 0,\\
\nn a_2(w) c_1(u) d_1(u\!+\!w) \!-\! a_2(u) a_2(u\!+\!w) d_1(w)
\!+\! b_2(u) b_2(u\!+\!w) d_1(w)\!-\!a_1(w) c_1(u\!+\!w) d_1(u) =
0. \\ \label{ybez1} \ena

Let us compare the third equation in the set (\ref{ybez1}) with
the sixth equation, after interchanging the variables $u$ and $w$
in the last one. Then taking out one from the other, we come to
(assuming, that $c(u)\neq 0$)
\bea \frac{a_1(w) - a_2(w)}{b_2(w)} = \frac{a_1(u+w) -
a_1(u+w)}{b_2(u+w)}. \label{au1}
 \ena
Now we can interchange the variables $u$ and $w$ in the fourth
equation in (\ref{ybez1}) and remove it from the first equation.
As a result the following relation holds
\bea \frac{a_1(u) - a_2(u)}{b_1(u)} = \frac{a_1(u+w) -
a_1(u+w)}{b_1(u+w)}. \label{au2}
 \ena

So, the following two constraints are followed from the equations
(\ref{au1}, \ref{au2})
\bea \frac{a_1(u) - a_2(u)}{b_1(u)} = \bar{\Delta},\qquad
b_2(u)=b_0 b_1(u), \label{au3}
 \ena
where $\bar{\Delta}$ and $b_0$ are constants. From the first and
sixth equations it follows, that  $(-1 + b_0^2) b_1(w) d_0 d_1(u)
d_1(u+w) = 0$. When the elements of the $R$-matrix are not $0$,
then the relation $ b_0^2=1$ is true. Taking into account this,
from the eighth and eleventh equations, we find $b_1(u) d_1(u+w)
(-1 + b_0) b(w)\bar{\Delta} ) = 0$. If $\bar{\Delta}\neq 0$, i.e.
$a_1(u)\neq a_2(u)$, then
\bea b_0=1.\label{z1}\ena
The independent equations in (\ref{ybez1}) now are nine, as one
can neglect fourth, sixth and eleventh equations, which are
contained in the remaining ones. Now let us consider the following
four equations: the first equation in (\ref{ybez1}), the second
one, the difference of the ninth and tenth equations and the
difference of the seventh and twelfth ones. The consistency
condition for the solutions of these linear equations in respect
of $a_1(w),\;b_1(w),\;c_1(w),\; d_1(w)$ reads as
\bea\nn&\Big(a_1(u+w) b_1(u)\!-\! a_1(u)
b_1(u+w)\Big)\Big(a_1(u+w)b_1(u)\!+\! b_1(u+w)(a_1(u)\!+\! b_1(u)
\bar{\Delta})\Big)&\\&\times(-1 + a_1^2(u)+b_1^2(u)-d_0
d_1^2(u)+a_1(u) b_1(u)\bar{\Delta})=0.&\ena
The vanishing of the first or the second brackets brings to
constant solutions. The equality of the expression in the third
bracket to zero presents nontrivial constraint on the solutions
\bea a_1^2(u)+b_1^2(u) -c_1^2(u)-d_0 d_1^2(u)=-a_1(u)
b_1(u)\bar{\Delta}. \label{z2}\ena
%

\paragraph{Summary of this subsection.}
 The constraints, laid by the YBE (\ref{ybee}) on the
$R(u)$-matrix (\ref{rx}) are of the following form
\bea \nn
&c_1/c_2=1,&\\
&b_2/b_1=b_0 \quad\&\quad b_0^2=1,\qquad
d_2/d_1=d_0,&\label{ast12}\ena
\bea\nn
 &\quad\left\{a_1/a_2=1 \;\;\; \&\ba{cc}b_0=1,&\;\;a_1 b_1=c_1 d_1 x,\;
\;\;\; a_2a_1-c_1c_2+b_1b_2-d_1 d_2 =2a_1
b_1\Delta\\b_0=-1,&\;\;\; a_1 a_2 +b_1 b_2 - c_1 c_2 \!-\!d_1
d_2=0, \ea\right\}^* &\\& \mb{or}\;\qquad
\left\{b_0=1,\;\;\;a_2-a_1=\bar{\Delta} b_1 \;\;\;\&\;\;\;
a_2a_1-d_1 d_2=c_1c_2-b_1b_2\right\}^{**}.&\label{summarizeg}\ena

Here, as it was used before, by the parameters
$b_0,\;d_0,\;x,\;\Delta,\;\bar{\Delta}$  we have denoted the
constants. As we have noted, when the functions $d_i\neq 0$, then
the parameter $b_0$ is not arbitrary.

In the following sections we shall give a rather detailed analysis
of the obtained constraints and discuss the possible
parameterizations of the elements of $R$-matrices, beginning from
the well known $XYZ$-model case. The corresponding 1d spin-chain
Hamiltonian operators will be discussed as well.

\section{General parametrization of the solutions}
\addtocounter{section}{0}\setcounter{equation}{0}

The obtained relations on the matrix elements suggest that it is
possible to write down different equivalent parameterizations of
the $R$-matrices by means of the Jacobi elliptic functions (as it
is  well known for the XYZ model matrix $R_{XYZ}$) and
trigonometric functions \cite{Baxt,GRS,ENIA}.
%
%
%
This can be achieved by choosing  $c(u)=1$ (normalization) and
fixing the function $b(u)$. Then determining the other elements
from the obtained constraints and YBE one can find out the whole
solution.

\subsection{$R_{XYZ}$-matrix:  $a_1=a_2$ and $b_1=b_2$}

At first we analyze the equations (\ref{ybez}) with constraints
(\ref{cz10}) and (\ref{czg}), corresponding to the  case in ${
}^*$ (\ref{summarizeg}) for which $b_0=1$. For definiteness we
take $d_0=1$, as after finding the function $d(u)$, we can put
$d_1(u)=e^{\gamma/2}d(u)$ and $d_2(u)=e^{-\gamma/2}d(u)$.

Setting $w=-u$ in the two first equations of  (\ref{ybez}), we
find
\bea \frac{d(u)}{a(u)}=-\frac{d(-u)}{a(-u)},\qquad
\frac{b(u)}{c(u)}=-\frac{b(-u)}{c(-u)}.\label{ant}\ena
So, ${b(u)}/{c(u)}$ and ${d(u)}/{a(u)}$ are odd functions. The
third equation gives
\bea a(u)a(-u)-b(u)b(-u)-c(u)c(-u)+d(u)d(-u)=0,\ena
which, taking into account the previous relations (\ref{ant}), can
be rewritten as
\bea
([a(u)]^2-[d(u)]^2)\frac{d(-u)}{d(u)}=([b(u)]^2-[c(u)]^2)\frac{c(-u)}{c(u)}.\label{cz1g}
\ena
 Taking $c(u)=1$, from the relation $a(0)=c(0)$
 found in the previous section,  it follows $a(0)=1$. If
 there is a point $u_0$, where $a(u_0)=0$, then the constraint (\ref{cz10})
gives $d(u_0)=0$. Putting $u=u_0$ in the second equation of the
set (\ref{ybez})
 we can find
\be a(w+u_0)=-b(w)/b(u_0) \qquad \mbox{or} \qquad
a(w)=-b(w-u_0)/b(u_0).\ee
The relation (\ref{czg}) imposes $b(u_0)=\pm c(u_0)=\pm 1$. One
can fix
 $b(u)={\mbox{sn}[u,k]}/{\mbox{sn}[\lambda,k]}$, as the YB equations
 reading now as transformation equations of the function
$b(u)$ coincide with those of the Jacobi elliptic functions
\cite{Baxt}. To the same conclusion one can come analyzing the
differential equations for the function $b(u)$ which can be
obtained from YBE. So, there are two possibilities $u_0=\pm
\lambda$ and
$a(u)=-{\mbox{sn}[u-\lambda,k]}/{\mbox{sn}[\lambda,k]}$ or
$a(u)={\mbox{sn}[u+\lambda,k]}/{\mbox{sn}[\lambda,k]}$. Let us fix
$a(u)={\mbox{sn}[u+\lambda,k]}/{\mbox{sn}[\lambda,k]}$, as the
other case can be obtained from this one by transformations $u \to
-u$ or $\lambda\to -\lambda$.
%
%
The function $d(u)$ from the constraint (\ref{cz10}) takes the
form
\bea d(u)=\mathrm{constant}\; a(u)
b(u)=\mathrm{constant}\;{\mbox{sn}[u+\lambda,k]\mbox{sn}[u,k]}/({\mbox{sn}[\lambda,k]})^2.
\ena
From the  equation (\ref{cz1g}), after putting $u=-\lambda$, one
gets that the value of the above $\mathrm{constant}\!=\!\pm k
(\mbox{sn}[\lambda,k])^2$. The interchange $k\leftrightarrow -k$
leaves the Jacobi-elliptic functions invariant. For definiteness
we choose
\bea d(u)=k\;{\mbox{sn}[u+\lambda,k]\mbox{sn}[u,k]}.\ena

So, in general, the homogeneous $R$-matrix of the eight-vertex (or
$XYZ$) model can be parameterized by two model parameters, $k$ and
$ \lambda$, as follows
\bea\label{ybezi}
 R_{xyz}(u)\!=\!\!\left(\!\!\ba{cccc}
  \frac{\mbox{sn}[{u+\lambda},\;k]}{\mbox{sn}[\lambda,\;k]}&0&0&{
  e^{\gamma/2}k\;\mbox{sn}[{\lambda+u},k]\mbox{sn}[u,k]}\\
0&\frac{\mbox{sn}[u,\;k]}{\mbox{sn}[\lambda,\;k]}&1&0\\
0&1&\frac{\mbox{sn}[u,\;k]}{\mbox{sn}[\lambda,\;k]}&0\\
{e^{-\gamma/2}k\;\mbox{sn}[{\lambda+u},k]\mbox{sn}[u,k]}
&0&0&\frac{\mbox{sn}[{u+\lambda},\;k]}{\mbox{sn}[\lambda,\;k]}
\ea\!\!\right).
 \ena

Then the constant in (\ref{czg}) can be written as
$2\mbox{cn}[\lambda,k]\mbox{dn}[\lambda,k]$.

Any other parameterizations can be obtained by changing the
function $b(u)$, which can be regarded as a replacement of the
spectral parameter. Two different parameterizations by elliptic
functions can be connected one with another by the transformation
rules of the elliptic functions \cite{ENIA}.

 The
"free-fermionic" case, when constant in (\ref{czg}) is zero,
corresponds to the $XY$-model. It takes place, say when
$\lambda=K(k)$, where $K(k)$ is the complete elliptic integral of
the first kind. The corresponding $R_{XY}(u)$ matrix has the form
\bea\label{rxy}
 R_{xy}(u)\!=\!\!\left(\!\!\ba{cccc}
 \frac{\mbox{cn}[{\mbox{u}},\;\mbox{k}]}{\mbox{dn}[\mbox{u},\;\mbox{k}]}&0&0&{
  e^{\gamma/2}k\;\frac{\mbox{cn}[\mbox{u},\;\mbox{k}]\mbox{sn}[\mbox{u},\mbox{k}]}
  {\mbox{dn}[\mbox{u},\;\mbox{k}]}}\\
0&\mbox{sn}[{u},\;k]&1&0\\
0&1&\mbox{sn}[u,\;k]&0\\
{e^{-\gamma/2}k\;\frac{\mbox{cn}[\mbox{u},\;\mbox{k}]\mbox{sn}[\mbox{u},\mbox{k}]}
  {\mbox{dn}[\mbox{u},\;\mbox{k}]}}
&0&0&\frac{\mbox{cn}[\mbox{u},\;\mbox{k}]}
{\mbox{dn}[\mbox{u},\;\mbox{k}]} \ea\!\!\right).
 \ena

As we see, for the extended case of $R_{XYZ}$ with $a_1(u)=a_2(u)$
the only new parameter arises due to $d_0$. 
  Something essentially new arises for the inhomogeneous cases
($b_1(u)\neq b_2(u)$ or $a_1(u)\neq a_2(u)$).

\subsection{Free fermionic solution with $a_1=a_2 $ and  $b_1=-b_2$}

Here we would like to consider the case in ${ }^*$
(\ref{summarizeg}) for which $b_0=-1$. It is a free fermionic
case.

 There are two possible solutions for this case. One of
them is a rather trivial and demands $a(u)=\pm b(u)$. It gives
 \bea R(u)=c(u)\left(\ba{cccc}
e^{\alpha u}&0&0&e^{\gamma}\\
0&\pm e^{\alpha u}&1&0\\
0&1&\mp e^{\alpha u}&0\\
- e^{-\gamma}&0&0&e^{\alpha u}\\
\ea\right).\label{rz}\ena
$\alpha,\;\gamma$ are arbitrary numbers (see also Section 5).

The second solution can be constructed by this way. We suppose
$b(0)=0$. Here we can choose $c(u)=1$ as in the previous, $XYZ$,
case. From the independent YB equations
we can obtain $a(0)=1$. The constraint (\ref{cz20x}) gives
$d(0)=0$ (and also $a'(0)=0$). Expansion of the first equation in
(\ref{ybez}) near the point $w=0$ brings to a simple differential
equation for function $d(u)$, solution of which is
\bea d(u)=\tan{[\alpha u]}/\sqrt{d_0}. \ena
From the next two equations, using $a'(0)=0$, we get
$a(u)+b(u)=e^{(\varepsilon u)}\sec{[\alpha u]}$ ($\alpha$ and
$\varepsilon$ are arbitrary numbers). Then it is easy to find that
$b(u)=\sinh{[\varepsilon u]}\sec{[\alpha u]}$ and
$a(u)=\cosh{[\varepsilon u]}\sec{[\alpha u]}$.

So, the matrix form of this solution looks like as follows

\bea\label{ybezi1}
 \tilde{R}(u)\!=\!\!\left(\!\!\ba{cccc}
 \cosh{[\varepsilon u]}\sec{[\alpha u]}&0&0&{
  e^{\gamma/2}\;\tan{[\alpha u]}}\\
0&\sinh{[\varepsilon u]}\sec{[\alpha u]}&1&0\\
0&1&-\sinh{[\varepsilon u]}\sec{[\alpha u]}&0\\
{e^{-\gamma/2}\tan{[\alpha u]}} &0&0&\cosh{[\varepsilon
u]}\sec{[\alpha u]}\ea\!\!\right).
 \ena
We have replaced $d_0$ by $e^\gamma$.

As the numbers $\alpha $ and $\varepsilon$ are arbitrary, we can
take the matrix $\tilde{R}(u)$ (\ref{ybezi1}) as a matrix which
has two spectral parameters $\tilde{R}(u;v)$ (after multiplication
by $\cos[v]$)
\bea\label{ybezuv}
 \tilde{R}(u;v)\!=\!\!\left(\!\!\ba{cccc}
 \cosh{[u]}&0&0&{
  e^{\gamma/2}\;\sin{[v]}}\\
0&\sinh{[u]}&\cos{[v]}&0\\
0&\cos{[v]}&-\sinh{[u]}&0\\
{e^{-\gamma/2}\sin{[v]}} &0&0&\cosh{[u]}\ea\!\!\right).
 \ena
 and satisfies to the following YB equations
\bea \sum_{j_1,j_2,j_3} R_{i_1 i_2}^{j_1 j_2}(u;v)R_{j_1 i_3}^{k_1
j_3}(u+w;v+y)R_{j_2 j_3}^{k_2
k_3}(w;y)=\qquad\qquad\qquad\nn\\\qquad\qquad\qquad\sum_{j_1,j_2,j_3}
R_{i_2 i_3}^{j_2 j_3}(w;y)R_{i_1 j_3}^{j_1 k_3}(u+w;v+y)R_{j_1
j_2}^{k_1 k_2}(u;v).\label{ybeuv}\ena
Note, that this free-fermionic solution is not followed from the
tree-parametric elliptic solutions presented in \cite{BS}.

\subsection{
The nonhomogeneous case with $a_1 \neq a_2$.}

Let us proceed further and write down the solutions for
the inhomogeneous case $a_1\neq a_2$. 
In the previous cases the solutions for the equations (\ref{ybez})
have been found, using the necessary conditions (${ }^\ast
\ref{summarizeg}$) and the properties of the Jakobi elliptic
functions \cite{Baxt}.
 In the
same way one can try to find appropriate parametrization for the
equations (\ref{ybez1}), taking into account the constraints
(\ref{au3}, \ref{z1}, \ref{z2}) (i.e. (${ }^{\ast\ast}
\ref{summarizeg}$)).

The analysis of the equations gives that $b_1(0)=0$,
$a_1(0)=c_1(0)$, $d_1(0)=0$ (elsewise we shall have constant
solutions). The first equation implies that $b_1(u)$ is an odd
function. Using it, from the second and fifth equations, placing
$w=-u$, we find $a_2(u)=a_1(-u)$, so
$$a_1(-u)=a_1(u)+\bar{\Delta} b_1(u).$$
 If there is a point $u_0$,
where $b(u_0)=1$, then we can use it, for parameterizing the
remaining functions. We can choose as in the previous cases
$b(u)={\mbox{sn}[u,k]}/{\mbox{sn}[u_0,k]}$.

The first equation in (\ref{ybez1}), after differentiating near
the point $w=0$, gives an expression for the function $a(u)\equiv
a_1(u)$ via the functions $b(u)$, $d(u)$ and their derivatives.
\bea  a(u)=\frac{1}{b'(0)}\left(
b(u)a'(0)+b'(u)+b(u)d(u)d'(0)\right),\qquad b''(0)=0.\ena

From the relation
\bea \label{con} a(u)a(-u)+b(u)^2-1-d(u)^2=0, \ena
 after expansion
it at the point $u=0$, we can find $a'(0)=-\bar{\Delta}b'(0)/2$.
Then placing the above expression of the function $a(u)$ into the
equation (\ref{con}), we find two possible solutions for the
function $d(u)$. From the same equation it follows also the
following relation
$$\bar{\Delta}^2-4+4 (\mbox{sn}[u_0,k])^2(1+k-d'(0)^2)=0.$$
As we can find from the analysis of the equations (seventh
equation in (\ref{ybez1})), $d(u)$ is an odd function, which gives
the following constraint
$$(1-d'(0)^2)(k^2-d'(0)^2)=0.$$
If we take $d'(0)^2=1$, then we come to the following
parametrization
\bea
a(u)=\frac{\mbox{dn}[u,k]}{\mbox{cn}[u,k]}-\frac{\bar{\Delta}\mbox{sn}[u,k]}{2\mbox{sn}[u_0,k]},\\
d(u)=\frac{\pm \mbox{dn}[u,k]\mbox{sn}[u,k]}{\mbox{cn}[u,k]},\\
\bar{\Delta}^2=4 (\mbox{dn}[u_0,k])^2. \ena
The corresponding $R(u)$-matrix looks like
{
\bea\label{rxydc}
 R_{1}(u,k)\!=\!\!\left(\!\!\!\ba{cccc}
\frac{\mbox{dn[u,k]}}{\mbox{cn[u,k]}}\pm\frac{\mbox{dn}[\mbox{u}_0,\mbox{k}]\mbox{sn[u,k]}}{\mbox{sn}[\mbox{u}_0,\mbox{k}]}
 &0&0& e^{\gamma/2}\frac{\mbox{dn[u,k]}\mbox{sn[u,k]}}{\mbox{cn[u,k]}} \\
0&\frac{\mbox{sn[u,k]}}{\mbox{sn}[\mbox{u}_0,\mbox{k}]}&1&0\\
0&1
&\frac{\mbox{sn[u,k]}}{\mbox{sn}[\mbox{u}_0,\mbox{k}]}&0\\
e^{-\gamma/2}\frac{\mbox{dn[u,k]}\mbox{sn[u,k]}}{\mbox{cn[u,k]}}&0&0&
\frac{\mbox{dn[u,k]}}{\mbox{cn[u,k]}}\mp\frac{\mbox{dn}[\mbox{u}_0,\mbox{k}]\mbox{sn[u,k]}}{\mbox{sn}[\mbox{u}_0,\mbox{k}]}
\ea\!\!\!\right)\!.
 \ena}
 When we choose
$d'(0)^2=k^2$, then
\bea
a(u)=\frac{\mbox{cn}[u,k]}{\mbox{dn}[u,k]}-\frac{\bar{\Delta}\mbox{sn}[u,k]}{2\mbox{sn}[u_0,k]},\\
d(u)=\frac{\pm k \mbox{cn}[u,k]\mbox{sn}[u,k]}{\mbox{dn}[u,k]},\\
\bar{\Delta}^2=4(\mbox{cn}[u_0,k])^2. \ena
{
\bea\label{rxycd}
 R_{2}(u,k)\!=\!\!\left(\!\!\!\ba{cccc}
\frac{\mbox{cn[u,k]}}{\mbox{dn[u,k]}}\pm\frac{\mbox{cn}[\mbox{u}_0,\mbox{k}]\mbox{sn[u,k]}}{\mbox{sn}[\mbox{u}_0,\mbox{k}]}
 &0&0& e^{\gamma/2}k\frac{\mbox{cn[u,k]}\mbox{sn[u,k]}}{\mbox{dn[u,k]}} \\
0&\frac{\mbox{sn[u,k]}}{\mbox{sn}[\mbox{u}_0,\mbox{k}]}&1&0\\
0&1
&\frac{\mbox{sn[u,k]}}{\mbox{sn}[\mbox{u}_0,\mbox{k}]}&0\\
e^{-\gamma/2}k\frac{
\mbox{cn[u,k]}\mbox{sn[u,k]}}{\mbox{dn[u,k]}}&0&0&
\frac{\mbox{cn[u,k]}}{\mbox{dn[u,k]}}\mp\frac{\mbox{cn}[\mbox{u}_0,\mbox{k}]\mbox{sn[u,k]}}{\mbox{sn}[\mbox{u}_0,\mbox{k}]}
\ea\!\!\!\right)\!.
 \ena}

These two solutions are connected one with another by the so
called "transformations of the first degree" of the elliptic
functions: $k\to 1/k,\; u\to k u$ (see Table $21.6-8$ in
\cite{Korn}).

 If $u_0=K(k)$, where $K$ is the complete elliptic
integral of the first kind with the module $k$, then the first
version $R_1$ corresponds exactly to the $R$-matrix, which we have
found in \cite{ShAS} for $2d$ IM. The second solution $R_2$ at
$u_0=K(k)$ corresponds to the $XY$-model matrix $R_{XY}$
(\ref{rxy}). Also, one can check, that the three-parametric
solution $R(u;\psi_1,\psi_2)$ in \cite{BS} coincides with
(\ref{rxycd}) up to renormalization of the $R$-matrix by an
overall function (i.e. by function $c(u;\psi_1,\psi_2)$) when the
following relations have been fulfilled $\psi_1=\psi_2=u_0$, $u\to
2u$.

All the obtained $R(u)$ matrices,  (\ref{ybezi}), (\ref{ybezi1}),
(\ref{rxydc}) and (\ref{rxycd}), after an appropriate
normalization (multiplication by a function) have a property of
unitarity
\bea
 R(u)[R(-u)]^{+}=I.
 \ena

The discussed inhomogeneous matrices (\ref{rz}), (\ref{ybezi1}),
(\ref{rxydc}) and (\ref{rxycd}) have the "free-fermionic"
property: $a_1 a_2+b_1 b_2=c_1 c_2 +d_1 d_2$. A solution, which
has no such property, mets amongst the exceptional cases,
discussed in the Section 5 (\ref{rff}). The symmetry (quite
similar to the case of the $sl_q(2)$ symmetry of the $XXZ$ model)
of the inhomogeneous trigonometric solutions with the
"free-fermionic" property  is unveiled in the article \cite{shkh}.

\section{The corresponding quantum one dimensional Hamiltonian
operators}
\addtocounter{section}{0}\setcounter{equation}{0}\setcounter{equation}{0}

In the article \cite{ShAS} the authors have presented the
$R$-matrix for two-dimensional IM \cite{Ons}. The main idea there
was to find the matrix which satisfies YBE just from the Boltzmann
weights of the 2d IM with coupling constants $J_1,\; J_2$. Then
using the Baxter's transformation \cite{Baxt}
\bea e^{\pm 2J_1}=cn(i \;u,k)\pm i \;sn(i\;u,k),\\
 e^{\pm 2J_2}=i (dn(i \;u,k)\pm 1)/(k \;sn(i\;u,k)).
 \ena
 and  after some unitary transformation,
the $4\times 4$-matrix, corresponding to the Boltzmann weight of
the 2d IM, takes the form
{\footnotesize \bea\label{r}
 R(u,k)\!=\!\left(\!\ba{cccc}
 \mathrm{cn}[i\; u,\;k]\;\mathrm{dn}[i\;u,\;k]-i\; k\; \mathrm{sn}[i\;u,k]
 &0&0&i\;\mathrm{sn}[i\;u,k]\mathrm{dn}[i\;u,\;k] \\
0&-i\; \mathrm{sn}[i\;u,k]&\mathrm{cn}[i\;u,\;k]&0\\
0&\mathrm{cn}[i\;u,\;k]
&-i\; \mathrm{sn}[i\;u,k]&0\\
i\;\mathrm{sn}[i\;u,k]\mathrm{dn}[i\;u,\;k]&0&0&  \mathrm{cn}[i\;
u,\;k]\;\mathrm{dn}[i\;u,\;k]+i\; k\; \mathrm{sn}[i\;u,k]
\ea\!\!\!\right)\!.
 \ena}
Note that \be R(0,k)=P,\ee where $P$ is the permutation matrix. We
have verified in \cite{ShAS} that ${R}(u,k)$ satisfies the
Yang-Baxter equation.

If one performs the transformation called by Akhiezer as the
"second main transformation of the first degree" \cite{ENIA}
(also, see \cite{Korn}):
\bea \mathrm{sn}(i\;u,k)=i
\frac{\mathrm{sn}(u,k')}{\mathrm{cn}(u,k')},\quad
\mathrm{cn}(i\;u,k)=\frac{1}{\mathrm{cn}(u,k')},\quad
\mathrm{dn}(i\;u,k)=\frac{\mathrm{dn}(u,k')}{\mathrm{cn}(u,k')},
\ena
in (\ref{r}), then it becomes equivalent to the matrix
(\ref{rxydc}) with $u_0=K$.

The Hamiltonian operators of the one-dimensional quantum
spin-chain models, corresponding to this $R$-matrix and the matrix
of the XY-model, which also
 describes free fermions, are different.
 The "check" version of the matrix (\ref{r})
  $\check{R}=R P$ in the limit of small $u$-s  or %
\be J_1\sim J \Delta t,\qquad e^{-J_2}\sim h \Delta t, \qquad
\Delta t \ll 1 ,\label{trot} \ee
can be written in this operator form $\textbf{R}$ by using
$\sigma_i$-matrices
\bea \textbf{R}_{IM}=\frac{1}{h\Delta t}(1\otimes1+2\Delta t(J
\sigma_1\otimes\sigma_1 +h (1\otimes\sigma_z+\sigma_z\otimes 1))).
\ena

While the $\check{R}$-matrix of the XY-model has not the term $h
(1\otimes\sigma_z+\sigma_z\otimes 1)$ corresponding to the
interaction with transverse magnetic field, and has the following
expansion
\bea \textbf{R}_{XY}=(1\otimes1+u (\mathbf{J}_1
\sigma_1\otimes\sigma_1 +\mathbf{J}_2 \sigma_2\otimes\sigma_2)).
\ena
Therefore the first one describes the quantum 1d Ising model in
the transverse field $h$,
\bea H_I=\sum_i (J \sigma_1[i]\sigma_1[i+1] +h \sigma_z[i]),\ena
and the R-matrix of the eight-vertex model (\ref{ybezi}), with the
free-fermionic condition, (\ref{rxy}), describes the quantum 1d
XY-model.
\bea H_{XY}=\sum_i (J_1 \sigma_1[i]\sigma_1[i+1] +J_2
\sigma_2[i]\sigma_2[i+1]).\ena
%


What is then the meaning of the new parameter $u_0$? One can
verify, that this parameter characterizes the parameters $J_1,\;
J_2$ of the $XY$ model in the transverse field. The matrices
$R_1(u)$ and $R_2(u)$ near the point $u=0$ have the following form
(we take $\gamma=0$)
{\bea\label{rh}
 \check{R}_{u_0}=R_{1,2}(u,k)|_{u\to 0}P=I+u\left(\!\ba{cccc}
 \pm\frac{\bar{\Delta}}{2 \mbox{sn}[\mbox{u}_0,\mbox{k}]}
 &0&0&d'(0) \\
0&0&\frac{1}{\mbox{sn}[\mbox{u}_0,\mbox{k}]}&0\\
0&\frac{1}{\mbox{sn}[\mbox{u}_0,\mbox{k}]}
&0&0\\
d'(0)&0&0& \mp \frac{\bar{\Delta}}{2
\mbox{sn}[\mbox{u}_0,\mbox{k}]}\ea\!\right)\!.
 \ena}
In the operator form the expansion of the matrix $\check{R}_{u_0}$
is
\bea \mathbf{R}_{u_0}=1\otimes
1+u/2\left(\mathbf{J}_1\sigma_1\otimes
\sigma_1+\mathbf{J}_2\sigma_2\otimes\sigma_2+\mathbf{h}(\sigma_z\otimes
1+1\otimes \sigma_z)\right).\ena
Here $J_1$ and $J_2$ are characterized by parameter $u_0$,
$$\mathbf{J}_1=\frac{d'(0)\;
\mbox{sn}[u_0,k]+1}{\mbox{sn}[u_0,k]}, \quad
\mathbf{J}_2=\frac{1-d'(0)\;\mbox{sn}[u_0,k]}{\mbox{sn}[u_0,k]},
\quad \mathbf{h}=\pm \frac{\bar{\Delta}}{2\mbox{sn}[u_0,k]}. $$
 The
corresponding 1d quantum spin chain Hamiltonian has the form
 \bea
H_{XY}=\sum_i (\mathbf{J}_1 \sigma_1[i]\sigma_1[i+1] +\mathbf{J}_2
\sigma_2[i]\sigma_2[i+1]+ \mathbf{h } \sigma_3[i]). \ena
 In the physical aspects two cases of $R_1$ and $R_2$ are the
same, and give $XY$ model in the transverse field.

The $R$-matrices (\ref{ybezuv}) with different $u$ and $v$ for the
case $b_0=-1$ have the following algebraic formulation
\bea \tilde{\mathbf{R}}(u;v)=\frac{\cosh{[u]}}{2}(1\otimes
1+\sigma_3\otimes \sigma_3)+i \frac{\sinh{[u]}}{2}(\sigma_1\otimes
\sigma_2-\sigma_2\otimes \sigma_1)+\\\nn
\frac{\cos{[v]}}{2}(1\otimes 1-\sigma_3\otimes
\sigma_3)+\sin{[v]}\Big(e^{\gamma/2}\sigma^+\otimes
\sigma^++e^{-\gamma/2}\sigma^-\otimes\sigma^-\Big).\label{ruv}\ena
At $u=\pm i v$ and $\gamma=\pm i \pi$ we arrive at
$\tilde{\mathbf{R}}(iv,v)=\cos{[v]}1\otimes 1\pm
\sin{[v]}\sigma_1\otimes \sigma_2$ or
$\tilde{\mathbf{R}}(iv,v)=\cos{[v]}1\otimes 1+\pm
\sin{[v]}\sigma_2\otimes \sigma_1$, which coincide with the
$R$-matrix, derived from the propagator in  \cite{KS}.

Setting $u=-i \Theta v$ and $\gamma=  -2i \eta$, and using the
expansion of the operator (\ref{ruv}) at the point $v=0$, we shall
come to the following spin-chain Hamiltonian operator
\bea
 &\tilde{H}=1/2 \sum_i \Big(\cos{\eta}
\Big(\sigma_1[i]\sigma_1[i+1]-\sigma_2[i]\sigma_2[i+1]\Big)
+\sin{\eta}\Big(
\sigma_1[i]\sigma_2[i+1]+\sigma_2[i]\sigma_1[i+1]\Big)+&\nn \\
&\Theta\Big(\sigma_1[i]\sigma_2[i+1]-\sigma_2[i]\sigma_1[i+1]\Big)
\Big).& \ena

\section{Exceptional cases \label{exc}}

In the course of the previous analysis we have excluded some
cases, which imply the following conditions on the matrix elements
of the $R$-matrix:\\
 $h_i(u)=f_i(u)\;\;\;$
or $\;\;\;h_{i}(u)=0$, $\;\;\;h,\;f=a,\;b,\;c,\;d$, $\;\;\;i=1,2$.

\subsection{ $h(u)=f(u)$, $\;\;\;h,\;f=a,\;b,\;c,\;d$.}

In case of homogeneous matrix, when $h_1(u)=h_2(u)$,
$h=a,\;b,\;c,\;d$, one can find out some special spectral
parameter-dependent solutions of YBE, which are not included in
the parameterizations of the previous sections. We try to consider
all  possible solutions for the each discussed case. Among them,
sometimes, there can be solutions, which are the limiting cases of
the solutions discussed in the previous sections.

$\bullet$ If one take $a(u)=c(u)$, then from YBE the equations
$b(u)=\pm d(u)$ and
\bea \frac{b(u)}{a(u)}-\frac{b(u+w)}{a(u+w)}+\frac{b(w)}{a(w)}
\left(1-\frac{b(u)}{a(u)}\frac{b(u+w)}{a(u+w)}\right)=0\ena
will follow. The solution is unique: $\frac{b(u)}{a(u)}=\tanh{u}$.

The resulting matrix is (up to a normalization function)
{\footnotesize \bea r_{a=c}(u)=\left(\!\!\ba{cccc}
 1&0&0&\pm \tanh{u} \\
0&\tanh{u}&1&0\\
0&1&\tanh{u}&0\\
\pm \tanh{u} &0&0&1 \ea\!\!\right).\ena}
This $R$-matrix has the free-fermionic property and is the
particular case of the solutions (\ref{rxydc}, \ref{rxycd}) with
the parameters $u_0=K(k),\; k=1$.

The choice $a(u)=-c(u)$ brings to constant solutions.

 $\bullet$ Let
$a(u)=\pm b(u)$. From YBE the relation $d(u)=\pm c(u)$ follows,
which is enough for the matrix to satisfy the all equations. So
here, besides of the normalization function the solution contains
one arbitrary function $f(u)(=d(u)/c(u))$:
{\footnotesize \bea r_{a=\pm b}(u)=\left(\!\!\ba{cccc}
 1&0&0&\pm f(u)\\
0&\pm 1&f(u)&0\\
0&f(u)&\pm 1&0\\
\pm f(u)&0&0&1 \ea\!\!\right). \label{rff}\ena}
In this case the matrix has free-fermionic property only for the
constant function $f(u)=\pm 1$.

$\bullet$ The  choice $a(u)=\pm d(u)$ gives only constant
solutions: $c(u)=\pm a(u)$, $b(u)=\pm a(u)$ up to the
normalization function.
{\footnotesize \bea r_{a=\pm d}(u)=a(u)\left(\!\!\ba{cccc}
 1&0&0&\pm 1\\
0&\pm 1&\pm 1&0\\
0&\pm 1&\pm 1&0\\
\pm 1&0&0&1 \ea\!\!\right).\ena}
$\bullet$ The other cases bring either to  constant or already
obtained solutions.
\subsection{$h_i(u)=0$}

Here we permit (at least) the vanishing of one of the functions
$h_i(u)$, $h=a,\;b,\;c,\;d$. We are coming to the rather trivial
solutions, when $h=a,\;c$.

\paragraph{Constant solutions ($u$ is included only in the phases).}  Examples of the  constant
solutions are:  when $a_1(u)=0$ (below $\alpha,\;\gamma$ are
arbitrary constants),
{\footnotesize \bea r_{a_1=0}=\left(\!\!\ba{cccc}
 0&0&0& e^{\gamma}\\
0&\pm \sqrt{2}&\pm 1&0\\
0&\pm 1&\pm \sqrt{2}&0\\
e^{-\gamma}&0&0&2 \ea\!\!\right),\ena}
or the matrix, when $d_1=0$
{\footnotesize \bea r_{d_1=0}=\left(\!\!\ba{cccc}
 1&0&0& 0\\
0&0& 1&0\\
0&0&\pm 1&0\\
e^{\gamma}&0&0&0 \ea\!\!\right).\ena}
A constant solution, when $b_1(u)=b_2(0)=0$ is
{\footnotesize \bea r_{b_1=0}=\left(\!\!\ba{cccc}
 1&0&0& e^{\gamma\pm i \pi/4}/\sqrt{2}\\
0&0& e^{\pm i \pi/4}/\sqrt{2}&0\\
0&e^{\pm i \pi/4}/\sqrt{2}&0&0\\
e^{-\gamma\pm i \pi/4}/\sqrt{2}&0&0&\pm i \ea\!\!\right).\ena}
Some constant solutions with $c_1(u)=0$ (which have not too much
zero matrix elements) are
{\footnotesize \bea r_{c_i=0}=\left(\!\!\ba{cccc}
 1&0&0& e^{\gamma+\alpha u}\\
0&\pm 1& 0&0\\
0&0&\pm 1&0\\
0&0&0&a_0 \ea\!\!\right)|_{a_0^2=1}.\ena}
and
{\footnotesize \bea r_{c_1=0}=\left(\!\!\ba{cccc}
 1&0&0& e^{\gamma}\\
0&a_0& 0&0\\
0&1+a_0 a_b&-a_b&0\\
0&0&0&a_0 a_b \ea\!\!\right)|_{a_0^2=1,\; a_b^2=1}.\ena}

 Of course, one must remember that, all the matrices $R$ are
defined up to a multiplicative arbitrary function $f(u)$: $R(u)\to
f(u)R(u)$, if $R(u)$ is a solution of YBE, then the matrix
$f(u)R(u)$ also satisfies the equations.

 For the constant solutions of YBE one can see the papers
 \cite{Hieta,Hlavati}.
\subsubsection*{ Spectral parameter dependent solutions}

We below give all the such solutions, which have spectral
parameter dependence (not only by an  overall factor function or a
phase function), provided that $f_i(u)=0$, with some $f=a,b,c,d$.

\paragraph{Homogeneous matrices.}  A nontrivial
solution arises when $b_1(u)=b_2(0)=0$, which is rather similar to
the XXZ model's $R$-matrix
{\footnotesize \bea \label{rxyzs}r_{X_{-}XZ}(u)
=\left(\!\!\ba{cccc}
 \sin[u_0]&0&0& \sin[u]\\
0&0& \sin[u+u_0]&0\\
0&\sin[u+u_0]&0&0\\
\sin[u]&0&0&\sin[u_0] \ea\!\!\right).\ena}
At the point $u=0$ the check version of the matrix (\ref{rxyzs})
can be represented in following operator form,
\bea {\bf{r_{X_{-}XZ}}}=\sin[u_0]1\otimes1+u \left(\mathbf{J}
(\sigma_1\otimes\sigma_1-\sigma_2\otimes\sigma_2)-\Delta
\sigma_3\otimes\sigma_3\right). \ena
Here $\mathbf{J}=1$ and $\Delta=\cos[u_0]$. So, this $R$-matrix
describes a $XYZ$ model with $J_1=-J_2$.
\bea H_{X_{-}X Z}=\sum_i (\mathbf{J}
(\sigma_1[i]\otimes\sigma_1[i]-\sigma_2[i]\otimes\sigma_2[i])
-\Delta \sigma_z[i]\otimes\sigma_z[i]),\ena
One can see, that $\bar{R}=r_{X_{-}X Z}(u)$ and the $R$-matrix of
the $XXZ$ model can be connected one with other by this
transformation
$\bar{R}_{ij}^{i'j'}=R_{i\;\bar{j}}^{i'\;\bar{j}'}$, where
$\bar{i}=mod[i+1]2$, i.e. it interchanges the indexes $0$ and $1$.

Note that the solution (\ref{rxyzs}) overlaps with the
inhomogeneous solution (\ref{ybezuv}). At $u_0=\pi/2$ the matrix
(\ref{rxyzs}) coincides with the $\tilde{R}(0,u)$. An arbitrary
constant $e^{\gamma}$ can be included also in $r_{X_{-}X Z}(u)$:
$d_1\to e^{\gamma} d_1$, $d_2 \to e^{-\gamma} d_2$.

 \paragraph{ Inhomogeneous matrices.}

A rational solution is obtained from the condition $b_1=0$, if it
is assumed, that $b_2(u)\neq 0$ and instead of it the relations
$d_1(u)=d_2(u)=0$ take place.
{\footnotesize\bea \label{su}r_{b_1=d_i=0}(u) =\left(\!\!\ba{cccc}
1&0&0& 0\\
0&u & 1&0\\
0&1&0&0\\
0&0&0&1 \ea\!\!\right).\ena}
A simple solution one can find out when, $b_i=0$ and $d_i=0$,
 $i=1,2$,
{\footnotesize \bea r_{b_i=d_i=0} =\left(\!\!\ba{cccc}
 e^{p\; u}&0&0& 0\\
0&0& 1&0\\
0&1&0&0\\
0&0&0&e^{q\; u}\ea\!\!\right),\ena}
where $p$ and $q$ are arbitrary numbers. A particular case with
$q=\pm p$ of this matrix can be obtained from the solution
$R_{XXZ}$ (\ref{rxxzi}), dividing it onto $\sin[u_0]$ and taking
the limit $u_0\to \infty$.

The case $b_i=0$, $i=1,2$ and $d_i\neq 0$, when $a_1 \neq a_2$
leads to  elliptic solutions. Here the number of the independent
YB equations is four (we set $c_i=1$, $d_1=d_2\equiv d$):
\bea a_1(u+w)-a_1(u)a_1(w)-a_2(u+w)d(u)d(w)=0,\\\nn
a_2(u+w)-a_2(u)a_2(w)+a_1(u+w)d(u)d(w)=0,\\\nn
a_1(u)d(w)-a_2(u)d(u+w)+a_2(u+w)a_2(w)d(u)=0,\\
a_2(u)d(w)-a_1(u)d(u+w)+a_1(u+w)a_1(w)d(u)=0. \ena
Finding that $a_i(0)=1$ and $d(0)=0$, and expanding the equations
near the point $u=0$, it follows that the function $d(u)$ must
satisfy the following differential equation $d''(u)=2
(\bar{a}_0)^2 d(u)^3+2((\bar{a}_0)^2+2 a_0^2) d(u)$, with some
definite $a_0$ and $\bar{a}_0$. The solution of this equation is
expressed by the Jacobi's elliptic function: $d(u)= \bar{a}_0
\mathrm{sn}[u,k]$. Then the functions $a_1(u)$ and $a_2(u)$ are
given by the following relations
$$a_1(2u)=\frac{2 a_0 d(u)}{\bar{a}_0(1+d(u)^2)}
+\frac{d'(u)}{\bar{a}_0(1-d(u)^2)},\;\;\;a_2(2u)=\frac{-2 a_0
d(u)}{\bar{a}_0(1+d(u)^2)} +\frac{d'(u)}{\bar{a}_0(1-d(u)^2)}.$$
The values of the parameters $a_0$ and $\bar{a}_0$ are expressed
by the module $k$ as
$$\bar{a}_0=\pm i \sqrt{k} \;\;\;\&\;\;\; a_0=\pm i(k-1)/2 \quad
\textrm{and}\quad  \bar{a}_0=\pm \sqrt{k} \;\;\;\&\;\;\; a_0=\pm
i(k+1)/2.$$

These solutions can be obtained from the matrix (\ref{rxycd})
taking $u_0=i K',\; \mathrm{sn}[i K',k]=\infty$ and making the
Landen's transformation of the elliptic functions, $k\to
2\sqrt{k}/(1+k),\;\;\;u\to (1+k)u$ \cite{Baxt}.

\paragraph{$d_i(u)=0$.}
Trigonometric solutions, similar to the $XXZ$ model's $R$-matrix,
exist when the constraints $d_2(u)=0$, $d_1\neq 0$ take place.
From the equations in (\ref{ybee}), which contain only the
functions $c_1(u),\;c_2(u)$, one finds out $c_2(u)=c_1(u)e^{\alpha
u}$. After comparison of some equations in (\ref{ybez1}), one
comes to $b_2(u)=b_1(u) b_0$, $b_0$ being a constant (if
$c_i(u)\neq 0$). Then, in the same way as previously, we are
coming to the relation (\ref{aa}), i.e.
$\frac{a_2(u)-a_1(u)}{b(u)}=\bar{\Delta}$ is a constant. Now let
us consider two cases with zero or non-zero values of
$\bar{\Delta}$ separately.

\paragraph{ The case $\bar{\Delta}=0$.}

 Then the independent equations in the set (\ref{ybez1})
are the following ones (provided that $d_1\neq 0$).
{\footnotesize\bea\nn
 &a_1(u) b_1(u\!+\!w) c_1(w) \!-\!b_1(w) c_1(u) c_1(u\!+\!w)
\!-\! a_1(u\!+\!w) b_1(u) c_1(w)  = 0,&
\\\nn &a_1(u) a_1(w) c_1(u\!+\!w) \!-\! b_0 b_1(w) b_1(u) c_1(u\!+\!w) \!-\! a_1(u\!+\!w) c_1(u) c_1(w)  = 0,&\\\label{eqa}
& b_0 a_1(w)
 b_1(u\!+\!w) c_1(u) \!-\! b_0 a_1(u\!+\!w) b_1(w) c_1(u) \!-\!
b_0 b_1(u)
c_1(u\!+\!w) c_1(w)  = 0,&\\
\nn &a_1(w) c_1(u\!+\!w) d_1(u) \!-\! a_1(w) c_1(u) d_1(u\!+\!w) \!+\! a_1(u) a_1(u\!+\!w) d_1(w) \!-\! b_1(u) b_1(u\!+\!w) d_1(w) = 0,&\\
\nn &b_0 b_1(u\!+\!w) c_1(w) d_1(u) \!+\! a_1(u) b_1(w)
d_1(u\!+\!w) \!-\! b_0 a_1(w) b_1(u) d_1(u\!+\!w) \!-\!
b_1(u\!+\!w) c_1(u) d_1(w)
= 0,&\\
\nn &a_1(u\!+\!w) a_1(w) d_1(u) \!-\! b_1(u\!+\!w) b_1(w) d_1(u)
\!-\! a_1(u) c_1(w) d_1(u\!+\!w) \!+\! a_1(u) c_1(u\!+\!w) d_1(w)=
0.&\ena}
First three equations coincide with the homogeneous ones in
(\ref{ybeh}). So the relation (\ref{cx}) takes place and the
unique non-trivial solutions can be parameterized as the matrix
elements of the $XXZ$ model -
$a_1(u)=\sin{[u+u_0]},\;b_1(u)=\sin{[u]},\; c_1=\sin{[u_0]}$. The
analysis of the remaining three equations gives the following
solution for the next function,
$d_1(u)=\sin{[u+u_0]}\sin{[u]}e^{\varepsilon}$, where
$\varepsilon$ is a constant number. The matrix representation of
this solution is
{
\bea \label{eps}r_{XXZ/d}(u) =\left(\!\!\ba{cccc}
\sin{[u+u_0]}&0&0& \sin{[u+u_0]}\sin{[u]}e^{\varepsilon}\\
0&\sin{[u]}&\sin{[u_0]}&0\\
0&\sin{[u_0]}&\sin{[u]}&0\\
0&0&0&\sin{[u+u_0]} \ea\!\!\right).\ena}
The operator representation of the expansion of the check matrix
at the point $u=0$
 looks like as follows
\bea {\bf{r}_{XXZ/d}}=\sin[u_0]1\otimes1+u\left(
(\sigma_1\otimes\sigma_1+\sigma_2\otimes\sigma_2)-\Delta
\sigma_3\otimes\sigma_3)+e^{\varepsilon}
\sigma_{+}\otimes\sigma_{+}\right). \ena
Here $\Delta=\cos[u_0]$ and $\sigma_{\pm}=(\sigma_{1}\pm i
\sigma_{2})/2$. The corresponding one dimensional spin-Hamiltonian
is
\bea H_{XXZ/d}=\sum_i (
(\sigma_1[i]\sigma_1[i+1]+\sigma_2[i]\sigma_2[i+1]) -\Delta
\sigma_z[i]\sigma_z[i+1]+e^{\varepsilon}
\sigma_{+}[i]\sigma_{+}[i+1]).\ena

With the general parametrization of the $XXZ$-model
$a_1(u)=a_2(u)=\sin{[u+u_0]},\;\\ b_1(u)=\sqrt{b_0}\sin{[u]},\;
b_2(u)=\sin{[u]}/\sqrt{b_0};\; c_1=e^{\alpha\; u}\sin{[u_0]},\;
c_2(u)=e^{-\alpha\; u}\sin{[u_0]}$, one will find  easily that the
only solution to (\ref{eqa}) corresponds to the choice
$\alpha=0,\; b_0=1$, i.e. coincides with the solution given above.

But we can check, that there is also another solution for the
particular case $\cos{[u_0]} = 0$. Then it turns out $b_0 =-1$.
Corresponding $R$-matrix looks like ($\alpha$ is an arbitrary
number)

\bea \label{eps}\tilde{r}_{XX/d}(u) =\left(\!\!\ba{cccc}
\cos{[u]}&0&0& \sinh{[\alpha u]}e^{\varepsilon}\\
0&\pm i\sin{[u]}&e^{\alpha u}&0\\
0&e^{-\alpha u}&\mp i\sin{[u]}&0\\
0&0&0&\cos{[u]} \ea\!\!\right). \ena

\paragraph{The case $\bar{\Delta}\neq 0$.} The set of the independent
equations now are
{\footnotesize\bea\nn
 &a_1(u) b_1(u\!+\!w) c_1(w) \!-\!b_1(w) c_1(u) c_1(u\!+\!w)
\!-\! a_1(u\!+\!w) b_1(u) c_1(w)  = 0,&
\\\nn &a_1(u) a_1(w) c_1(u\!+\!w) \!-\! b_1(w) b_2(u) c_1(u\!+\!w) \!-\! a_1(u\!+\!w) c_1(u) c_1(w)  = 0,&\\\nn &a_1(w)
b_2(u\!+\!w) c_1(u) \!-\! a_1(u\!+\!w) b_2(w) c_1(u) \!-\! b_2(u) c_1(u\!+\!w) c_1(w)  = 0,&\\
\nn
&a_2(u\!+\!w) b_1(w) c_1(u) \!+\! b_1(u) c_1(u\!+\!w) c_1(w) \!-\!a_2(w) b_1(u\!+\!w) c_1(u) = 0,&\\
\nn & b_1(u) b_2(w) c_1(u\!+\!w) \!+\! a_2(u\!+\!w) c_1(u)
c_1(w)\!-\!a_2(u) a_2(w) c_1(u\!+\!w)  = 0,\\
 & a_2(u) b_2(u\!+\!w) c_1(w) \!-\!b_2(w) c_1(u) c_1(u\!+\!w) \!-\! a_2(u\!+\!w) b_2(u) c_1(w) = 0,& \label{eqa1}\\
\nn  & a_2(w) c_1(u\!+\!w) d_1(u) \!-\! a_1(w) c_1(u) d_1(u\!+\!w) \!+\! a_1(u) a_1(u\!+\!w) d_1(w) \!-\! b_1(u) b_1(u\!+\!w) d_1(w) = 0,&\\
\nn & b_2(u\!+\!w) c_1(w) d_1(u) \!+\! a_1(u) b_1(w) d_1(u\!+\!w)
\!-\! a_1(w) b_2(u) d_1(u\!+\!w) \!-\! b_1(u\!+\!w) c_1(u) d_1(w)
= 0,&\\ \nn  & b_2(u\!+\!w) b_2(w) d_1(u) \!+\! a_1(u) c_1(w) d_1(u\!+\!w) \!-\! a_2(u) c_1(u\!+\!w) d_1(w) \!-\!a_1(u\!+\!w) a_1(w) d_1(u)= 0,&\\
\nn & a_2(u\!+\!w) a_2(w) d_1(u) \!-\! b_1(u\!+\!w) b_1(w) d_1(u) \!-\! a_2(u) c_1(w) d_1(u\!+\!w) \!+\! a_1(u) c_1(u\!+\!w) d_1(w) = 0,&\\
\nn & a_2(w) b_1(u) d_1(u\!+\!w) \!-\! a_2(u) b_2(w) d_1(u\!+\!w) \!+\! b_2(u\!+\!w) c_1(u) d_1(w)\!-\!b_1(u\!+\!w) c_1(w) d_1(u)  = 0,&\\
\nn & a_2(w) c_1(u) d_1(u\!+\!w) \!-\! a_2(u) a_2(u\!+\!w) d_1(w)
\!+\! b_2(u) b_2(u\!+\!w) d_1(w)\!-\!a_1(w) c_1(u\!+\!w) d_1(u) =
0.&\ena}
The solutions of the first six equations are known. Let us take
most general parameterizations of them as
$a_1(u)=\sin{[u+u_0]},\;b_1(u)=\sqrt{b_0}\sin{[u]},\;
b_2(u)=\sin{[u]}/\sqrt{b_0};\; c_1=e^{\alpha\; u}\sin{[u_0]},\;
c_2(u)=e^{-\alpha\; u}\sin{[u_0]}, a_2(u)=\sin{[u_0-u]}$. Then
from the next two equations we are finding a relation $d(u)=
e^{\alpha
u}\mathrm{csc}[u_0]\sin{[u]}d'(0)(\sin{[u+u_0]}+b_0\sin{[u_0-u]})/(1+b_0)$.
The remaining equations are giving constraints on the values of
$\alpha$ and $b_0$. Finally we are arriving at the following
solutions. For $\alpha=0$ and $b_0=1$, the $R$-matrix is
\bea \label{epsi}r_{XYZ/d}(u) =\left(\!\!\ba{cccc}
\sin{[u+u_0]}&0&0& \cos{[u]}\sin{[u]}e^{\varepsilon}\\
0&\sin{[u]}&\sin{[u_0]}&0\\
0&\sin{[u_0]}&\sin{[u]}&0\\
0&0&0&\sin{[u_0-u]} \ea\!\!\right),\ena
and for the pairs $\alpha=i$ and $b_0=e^{2i u_0}$ and
$\alpha=-i,\;b_0=e^{-2i u_0}$ we find
 \bea
\label{epspm}r_{XYZ/d}^{\pm}(u) =\left(\!\!\ba{cccc}
\sin{[u+u_0]}&0&0& \sin{[u]}e^{\varepsilon}\\
0&\sin{[u]}e^{\pm i u_0}&\sin{[u_0]}e^{\pm iu}&0\\
0&\sin{[u_0]}e^{\mp iu}&\sin{[u]}e^{\mp i u_0}&0\\
0&0&0&\sin{[u_0-u]} \ea\!\!\right),\ena
where $\varepsilon$ is a constant, which is defined as
$e^{\varepsilon}=d'(0)$.

Of course, the transpositions of the matrices (\ref{epsi}) and
(\ref{epspm}) are also solutions of YBE.

The $R$-matrix in (\ref{epsi}) can be found from the matrix
(\ref{rxycd}) by taking
 an appropriate limit $r_{XYZ/d}(u)=\lim_{k\to 0,\;
 e^\gamma=e^\varepsilon/k}{R_2(u,k)}$. In contrast to this, the matrix
 (\ref{epspm}) is not included in the limiting set of the solutions
obtained in the previous sections.  Note, that the
solutions  (\ref{epspm}) and  (\ref{eps}) overlaps at the
particular values of the parameters $u_0$ and $\alpha$,  $u_0=\pm
\pi/2$ and $\alpha =\pm i$.

 The operator representation of the check matrix of (\ref{epsi}) at the point $u=0$
looks like
\bea {\bf{r}_{XYZ/d}}=\sin[u_0] 1\otimes1+u\left(
\sigma_1\otimes\sigma_1+\sigma_2\otimes\sigma_2+\Delta/2(\sigma_3\otimes
1+1\otimes \sigma_3)+e^{\varepsilon}
\sigma_{+}\otimes\sigma_{+}\right). \ena
Here, as in the previous case, $\Delta=\cos[u_0]$ and
$\sigma_{\pm}=(\sigma_{1}\pm i \sigma_{2})/2$. The corresponding
one dimensional spin-Hamiltonian is
\bea \label{h}H_{XYZ/d}=\sum_i (
\sigma_1[i]\sigma_1[i+1]+\sigma_2[i]\sigma_2[i+1]+
\Delta\sigma_3[i]+e^{\varepsilon}
\sigma_{+}[i]\sigma_{+}[i+1]).\ena

The one-dimensional spin-Hamiltonian appropriate for the
$R$-matrix (\ref{epspm}) contains some additional terms
%
\bea \label{hpm}H_{XYZ/d}^{\pm}=\sum_i \Big(
\sigma_1[i]\sigma_1[i+1]+\sigma_2[i]\sigma_2[i+1]\pm
i/2(\sigma_3[i]-\sigma_3[i+1])\qquad \\\nn +(e^{\pm i
u_0}-1)\sigma_+[i]\sigma_-[i+1]+(e^{\mp i
u_0}-1)\sigma_-[i]\sigma_+[i+1]+\Delta\sigma_3[i]+e^{\varepsilon}
\sigma_{+}[i]\sigma_{+}[i+1]\Big).\ena
\section
{Tetrahedral algebra of Zamolodchikov}
 \setcounter{equation}{0}

 There are very few solutions to the Zamolodchikovs' tetrahedral
 equations (ZTE) \cite{Zt,Korep}. 
  Their vertex versions
 can be served as the analogs of the vertex Yang-Baxter equations for three
 dimensional case.
 The solutions to ZTE, as well as the solutions to the equations presented in \cite{SSh}
  ensure the integrability of the models constructed therein.
   One of established ways of looking for the solutions of ZTE is connected
 with the use of solutions of YBE \cite{Korep,HSH}. The brief description of it
 is the following.  
If $R^{i}_{k\; p}(u,v),\; i=1,2$, are matrices acting
non-trivially on the space $V_k \otimes V_k$, and have following
properties
\bea R^{1}_{12}(u,v)R^{1}_{13}(u,w)R^{1}_{23}(v,w)
=R^{1}_{23}(v,w)R^{1}_{13}(u,w)R^{1}_{12}(u,v),\label{rrs}\\
 R^{1}_{12}(u,v)R^{2}_{13}(u,w)R^{2}_{23}(v,w)
=R^{2}_{23}(v,w)R^{2}_{13}(u,w)R^{1}_{12}(u,v),\label{rri} \ena
then the matrix $W_{i_1 j_1 k_1}^{i_2 j_2 k_2}$, obtained from
the relation, known as tetrahedral Zamolodchikov's algebra 
\bea  R^{i_1}_{12}(u,v)R^{j_1}_{13}(u,w)R^{k_1}_{23}(v,w)\!
=\!\!\!\sum_{i_2 j_2 k_2}\!\!\!W(u,v,w)_{i_1 j_1 k_1}^{i_2 j_2
k_2}(u,v,w) R^{k_2}_{23}(v,w)R^{j_2}_{13}(u,w)R^{i_2}_{12}(u,v)\nn\\
\label{kpw}\ena
can be served as a good candidate to the solution of the vertex
version of ZTE
\bea W_{i_1 i_2 i_3}^{j_1 j_2 j_3}(u_1,u_2,u_3)W_{j_1 i_4
i_5}^{k_1 j_4 j_5}(u_1,u_2,u_4)W_{j_2 j_4 i_6}^{k_2 k_4
j_6}(u_1,u_3,u_4)W_{j_3 j_5 j_6}^{k_3 k_5 k_6}(u_2,u_3,u_4)\\\nn
=W_{i_3 i_5 i_6}^{j_3 j_5 j_6}(u_2,u_3,u_4)W_{i_2 i_4 j_6}^{j_2
j_4 k_6}(u_1,u_3,u_4)W_{i_1 j_4 j_5}^{j_1 k_4
k_5}(u_1,u_2,u_4)W_{j_1 j_2 j_3}^{k_1 k_2 k_3}(u_1,u_2,u_3).\ena
At least this relation will be hold on the space
$$R^{k_1}_{34}(u_3,u_4)R^{k_2}_{24}(u_2,u_4)R^{k_3}_{14}(u_1,u_4)
R^{k_4}_{2,3}(u_2,u_3)R^{k_5}_{1,3}(u_1,u_3)R^{k_6}_{12}(u_1,u_2)\;|v_1\rangle|v_2\rangle|v_3\rangle|v_4\rangle,$$
which can be easily verified repeatedly using the equality
(\ref{kpw}). The written product of the $R_{ij}$-matrices is
defined on the tensor product of two-dimensional spaces
$\Big(V_1\otimes V_2 \otimes V_3 \otimes V_4\Big)$.

For the  matrix $R^{1}(u,v)$ coinciding with the homogeneous
$XY$-model's $R_{XY}(u-v)$,
 the corresponding $R^{2}(u,v)$ and
$W(u,v,w)$-matrices are found \cite{Korep,HSH}
 and it is proven, that $W(u,v,w)$
is a solution of ZTE with $R^{2}(u,v)$ chosen as
\be R^{2}(u,v)_{i_1 j_1}^{i_2 j_2}=(-1)^{j_1+1}R^{1}(u+v)_{i_1
j_1}^{i_2 j_2}. \label{r12}\ee
 In the same time  the following equations take place
\bea R^{2}_{12}(u,v)R^{1}_{13}(u,w)R^{1}_{23}(v,w)
=R^{1}_{23}(v,w)R^{1}_{13}(u,w)R^{2 \tau}_{12}(u,v),\ena
where
$$R^{2 \tau}(u,v)_{i_1 j_1}^{i_2 j_2}=(-1)^{i_1+1}R^{1}(u+v)_{i_1 j_1}^{i_2 j_2}.$$

Choosing the matrix $R^1_{ij}$ as one of the inhomogeneous
$R_{ij}$-matrices, discussed in this paper, and taking it's pair
$R^2_{ij}$ in the same way as in (\ref{r12}), we come to the
conclusion, that the $W(u,v,w)$-matrices satisfying to (\ref{kpw})
exist for the particular cases which coincide or are equivalent to
the homogeneous cases \cite{Korep,HSH} after some automorphisms
(mainly it brings to the change of the signs of some matrix
elements of $W(u,v,w)$). In case of the general inhomogeneous
matrices, there are no such $W(u,v,w)$-matrices.

Here we present another way of choosing of the pair $R^{i}$, for
which there is a $W(u,v,w)$-matrix satisfying Eq.(\ref{kpw}).
However this matrix is not a solution of ZTE, but it is an
example, that it is possible to construct the algebra (\ref{kpw})
by choosing $R^{i}_{ij}(u,v)$-matrices in various ways. By direct
calculations it can be verified, that the matrices
\bea R^{1}(u_1,u_2)=\left(\ba{cccc}\cos{[u_1-u_2]}&0&0&0\\
0&i\sin{[u_1-u_2]}&1&0\\
0&1&-i\sin{[u_1-u_2]}&0\\0&0&0&\cos{[u_1-u_2]}\ea\right),\\
R^2(u_1,u_2)=\left(\ba{cccc}1&0&0&\sin{[u_1-u_2]}\\0&0&\cos{[u_1-u_2]}&0\\
0&\cos{[u_1-u_2]}&0&0\\\sin{[u_1-u_2]}&0&0&1\ea\right), \ena
which are particular cases of the matrix (\ref{ybezuv}),
 both are the solutions of ordinary YBE, but for them the
relation (\ref{rri}) does not take place. The following (unique)
$W(u,v,w)$-matrix ensures the validity of the tetrahedral
Zamolodchikov algebra.
\bea & W_{000}^{000}=W_{111}^{111}=1,\qquad\nn&\\&\nn
W_{ijk}^{i'j'k'}={\bar{W}}_{ijk}^{i'j'k'}\left(2
f[u_1,u_2,u_3]-\sin{[2(u_1-u_2)]}\sin{[2(u_2-u_3)]}\right)^{-1},&\\&
f[u_1,u_2,u_3]=\sin{[u_1-u_2]}^2+\sin{[u_2-u_3]}^2+\sin{[u_1-u_3]}^2,&\\&\nn
{\bar{W}}_{001}^{001}={2
f[u_1,u_2,u_3]\cos{[u_2-u_3]}\csc{[u_1-u_3]}\sin{[u_1-u_2]}},&\\&\nn
{\bar{W}}_{010}^{010}= {-\sin{[2(u_1-u_2)]}\sin{[2(u_2-u_3)]}},
\quad {\bar{W}}_{001}^{010}=
{\cos{[u_1-u_2]}\sin{[u_2-u_3]}\sin{[u_1-u_3]}},&\\&\nn
{\bar{W}}_{010}^{001}= {2
f[u_1,u_2,u_3]\csc{[u_1-u_3]}^2\sin{[u_2-u_3]}^2},\quad
\nn{\bar{W}}_{000}^{011}= {-4
\sin{[u_1-u_2]}^2\sin{[u_2-u_3]}^2},&\\& \nn
{\bar{W}}_{011}^{011}={2
f[u_1,u_2,u_3]\cos{[u_1-u_2]}\csc{[u_1-u_3]}\sin{[u_2-u_3]}},&\\&\nn
{\bar{W}}_{100}^{001}= {-4
\cos{[u_2-u_3]}\csc{[u_1-u_3]}\sin{[u_1-u_2]}\sin{[u_2-u_3]}^2},&\\&\nn
{\bar{W}}_{100}^{010}= {4
\cos{[u_2-u_3]}\sin{[u_1-u_2]}\sin{[u_1-u_3]}},\quad
{\bar{W}}_{111}^{001}= {-4
\csc{[u_1-u_3]}^2\sin{[u_2-u_3]}^2},&\\&\nn {\bar{W}}_{111}^{010}=
{4},\quad {\bar{W}}_{101}^{011}= {2 f[u_1,u_2,u_3]},\quad
{\bar{W}}_{110}^{011}= {-4
\cos{[u_1-u_2]}\csc{[u_1-u_3]}\sin{[u_2-u_3]}^2},&\\&\nn
{\bar{W}}_{001}^{100}= {-4
\cos{[u_1-u_2]}\csc{[u_1-u_3]}\sin{[u_1-u_2]}^2\sin{[u_2-u_3]}},
&\\&\nn {\bar{W}}_{000}^{101}= {4
\sin{[u_1-u_2]}\sin{[u_2-u_3]}^2},\quad {\bar{W}}_{010}^{100}= {2
f[u_1,u_2,u_3]\csc{[u_1-u_3]}^2\sin{[u_1-u_2]}^2},&\\&\nn
{\bar{W}}_{011}^{101}= {4
\cos{[u_2-u_3]}\csc{[u_1-u_3]}\sin{[u_1-u_2]}^3},\quad{\bar{W}}_{000}^{110}=
{-4\sin{[u_1-u_2]}^2\sin{[u_2-u_3]}^2},&\\&\nn
{\bar{W}}_{011}^{110}={-4
\cos{[u_2-u_3]}\csc{[u_1-u_3]}\sin{[u_1-u_2]}^3},\quad
{\bar{W}}_{111}^{100}={-4\csc{[u_1-u_3]}^2\sin{[u_1-u_2]}^2},&\\&\nn
{\bar{W}}_{100}^{100}= {-2
f[u_1,u_2,u_3]\cos{[u_1-u_2]}\csc{[u_1-u_3]}\sin{[u_2-u_3]}},&\\&\nn
{\bar{W}}_{101}^{101}=
{-\sin{[2(u_1-u_2)]}\sin{[2(u_2-u_3)]}},\quad
{\bar{W}}_{110}^{101}={4
\cos{[u_1-u_2]}\csc{[u_1-u_3]}\sin{[u_2-u_3]}},&\\&\nn
{\bar{W}}_{101}^{110}={2 f[u_1,u_2,u_3]},\quad
{\bar{W}}_{110}^{110}= {2
f[u_1,u_2,u_3]\cos{[u_2-u_3]}\csc{[u_1-u_3]}\sin{[u_1-u_2]}}.&\\&
&\ena

\section{Conclusions}
\setcounter{equation}{0}

In this article we give all possible solutions of the YBE with
general inhomogeneous spectral parameter dependent $R(u)$-matrix
corresponding to the six-and eight-vertex models. The symmetry
relations imposed by the Yang-Baxter equations on the elements of
the general inhomogeneous $R$-matrices, together with the
consistency conditions are obtained. Thus, we present more
complete
classification of the solutions, than it was done before. 
The main conclusion about the nature of the solutions is, that
besides of the known homogeneous solutions (corresponding to the
$XXZ$ and $XYZ$ models), which admit redefinitions (including
parameters $\alpha, \; b_0, \; d_0$) due to gauge transformations,
the all other solutions - inhomogeneous or homogeneous, with the
behavior $\check{R}(0)\approx I$ (important in the context of the
integrable theory), have the
 "free-fermionic" property. 
  As it is known, this property ensures, that the corresponding
  physical models are exactly solvable \cite{WuChan, ShAS}.
 The fact that the "free-fermionic" $R$-matrices
  satisfy to the Yang-Baxter equations,
  hints that they admit some underlying symmetry \cite{GRS,shkh}.
    It is remarkable, that
among the exceptional solutions discussed in the Section 5  we met
such spectral-parameter dependent solutions (\ref{rff}, \ref{eps},
\ref{epspm}) which are not limit cases of the solutions with
general structure. The one-parametric solution (\ref{ybezi1}),
obtained in the subsection 3.2 gives rise to the two-parametric
solution (\ref{ybezuv}) to the YBE (\ref{ybeuv}). In the Section 6
we discuss the possibility to use the obtained matrices for the
construction of the solutions to the Zamolodchikov's Tetrahedral
Algebra. All the quantum chain-models constructed with the
obtained inhomogeneous $R(u)$-matrices ($R_{00}^{00}\neq
R_{11}^{11}$) describe spin-$1/2$ models with nearest-neighbor
interactions in a transverse magnetic field. By means of the
spin-fermion transformations, the corresponding Hamiltonian
operators describe nearest-neighbor interactions of free spin-less
fermions on a chain.

\paragraph{Acknowledgement} The work is partly supported by
Armenian Government grant 11-1c028.

\end{document}